\documentclass{aastex}

\slugcomment{}
\shorttitle{}
\shortauthors{Ikeda et al.}
\begin{document} 
\title{LARGE SCALE MAPPING OBSERVATIONS \\OF THE CI ($^{3}P_{1}-^{3}P_{0}$) 
AND CO ($J=3-2$) LINES\\ TOWARD THE ORION A MOLECULAR CLOUD}

\author{Masafumi Ikeda\altaffilmark{1}, Hiroyuki Maezawa\altaffilmark{1}, 
Tetsuya Ito\altaffilmark{1}, Gaku
 Saito\altaffilmark{1}, Yutaro Sekimoto\altaffilmark{1,2}, 
Satoshi Yamamoto\altaffilmark{1}, 
Ken'ichi Tatematsu\altaffilmark{3}, Yuji Arikawa\altaffilmark{2,4}, 
Yoshiyuki Aso\altaffilmark{2,5}, Takashi Noguchi\altaffilmark{3},
Sheng-Cai Shi\altaffilmark{3,6} , Keisuke
Miyazawa\altaffilmark{3}, 
Shuji Saito\altaffilmark{7}, Hiroyuki Ozeki\altaffilmark{7,8}, 
Hideo Fujiwara\altaffilmark{7,9}, 
Masatoshi Ohishi\altaffilmark{10}, and 
Junji Inatani\altaffilmark{11}}

\altaffiltext{1}{Research Center for the Early Universe and Department of
Physics, The University of Tokyo, Tokyo 113-0033, Japan.}
\altaffiltext{2}{Present Address : Nobeyama Radio Observatory,  
   National Astronomical Observatory of Japan,
    Nagano 384-1305, Japan.}
\altaffiltext{3}{ Nobeyama Radio Observatory,  
   National Astronomical Observatory of Japan,
   Nagano 384-1305, Japan.}  
\altaffiltext{4}{Department of Astronomical Science, the Graduate
University for Advanced Studies, Nobeyama Radio Observatory,
Nagano 384-1305, Japan.}
\altaffiltext{5}{Department of Astronomy, The University of Tokyo, 
Tokyo 113-0033, Japan.}
  \altaffiltext{6}{Present Address : Purple Mountain
   Observatory, Nanjing, JiangSu 210008, China.}
\altaffiltext{7}{Institute for Molecular Science,
 Okazaki 444-8585, Japan.}
  \altaffiltext{8}{Present Address : National Space Development
Agency of Japan, Tsukuba, Ibaraki 305-8505, Japan.}
  \altaffiltext{9}{Present Address : Department of Chemical System Engineering, 
School of Engineering, The University of Tokyo, 
 Tokyo 113-0033, Japan.}
\altaffiltext{10}{National Astronomy Observatory of Japan, Mitaka, Tokyo 
181-8588, Japan.}
\altaffiltext{11}{National Space Development
Agency of Japan, Tsukuba, Ibaraki 305-8505, Japan.}

\begin{abstract}
Large scale mapping observations of the $^{3}P_{1}$--$^{3}P_{0}$ fine structure 
transition of atomic carbon (CI, 492 GHz) and 
the $J$=3-2 transition of CO (346 GHz) toward the Orion A molecular cloud 
have been carried out with the Mt. Fuji submillimeter-wave telescope. 
The observations cover 9 square degrees, and include the Orion nebula M42 and 
the L1641 dark cloud complex. The CI emission extends over almost 
the entire region of the Orion A cloud and is surprisingly similar to that of $^{13}$CO($J$=1-0). 
The CO($J$=3-2) emission shows a more 
featureless and extended distribution than CI.
The CI/CO($J$=3-2) integrated intensity ratio shows a spatial gradient 
running from the north (0.10) to the south (1.2) of the Orion A cloud, 
which we interpret as a consequence of the temperature gradient. On the other hand, 
the CI/$^{13}$CO($J$=1-0) intensity ratio shows no systematic gradient. 
We have found a good correlation between the CI and $^{13}$CO($J$=1-0) intensities 
over the Orion A cloud. 
This result is discussed on the basis of photodissociation region models.

----------------------------------------------
  
\end{abstract}


\section{Introduction}

Neutral atomic carbon (CI) plays important roles in cooling and 
chemical processes in interstellar clouds, and its submillimeter-wave transitions 
($^{3}P_{1}$--$^{3}P_{0}$, 492 GHz; $^{3}P_{2}$--$^{3}P_{1}$, 809 GHz) have been observed toward various 
objects. The detailed distribution of CI around 
representative objects including photodissociation regions (PDR) has been studied 
at high angular resolution (e.g., Minchin et al. 1994; Tauber et al. 1995; White and Sandell 1995). 
Since CI is widely distributed throughout our Galaxy according to the data from 
the COBE satellite \citep{wr91}, it is of fundamental importance to map its 
large scale distribution over molecular clouds. Pioneering studies in 
this direction have been made using a focal reducer installed on the CSO 10 m 
antenna \citep{pl94,pl99,tate99}, 
to survey the distribution of CI toward several molecular clouds with a moderate resolution of $\sim3'$. 
In spite of these efforts, the observed areas are still limited relative to available  
maps of CO and its isotopomers. With this in mind, we have recently constructed a 1.2 m 
submillimeter-wave telescope for the exclusive use of CI survey 
observations at the summit of Mt. Fuji. 

The Orion A cloud is the nearest giant molecular cloud, and is located at about 450 pc 
from the Sun \citep{ge89}. Extensive observations of 
the cloud have been made in CO ($J$ = 2-1) \citep{sak94}, $^{13}$CO ($J$ = 1-0) 
\citep{bal87,nag98}, CS ($J$ = 1-0) \citep{tate93} and CS ($J$ = 2-1) \citep{tate98}. 
These observations have revealed that numerous dense cores, some of which are the birthplaces of new 
stars, are distributed throughout the cloud. The northern part of the Orion A 
cloud is known to be an active site of massive star formation. As a result, the 
cloud is illuminated by strong UV radiation from OB stars ($G_0$ is $\sim$$10^5$ in the 
vicinity of Orion KL). By contrast, the 
central and southern parts of the Orion A cloud are more quiescent, and known as 
the L1641 dark cloud. Although a number of low mass protostars, 
T Tauri stars, and H$\alpha$ emission-line stars are present, no massive stars are 
found there, and the UV radiation field is much weaker ($G_0\sim$1-5). 
Therefore, the Orion A cloud is a good target for studying structure of a molecular 
cloud under various UV field strengths.

In contrast to the extensive studies of the molecular gas distribution, only a few mapping 
observations of CI have been reported toward small portions of the Orion A cloud. 
\citet{wh91} and \citet{wh95} observed the Orion-KL region with a $9''.8$ beam. 
\citet{tau95} reported a 15$''$ map of the Orion bright bar and Orion-S cloud. 
\citet{tate99} explored the CI distribution in a part of the $\int$-shaped filament 
with a focal reducer system on the CSO. In this paper, we present the first large scale maps 
of CI and CO($J$=3-2) covering the entire region of the Orion A cloud.

\section{Observations}

The CI($^{3}P_{1}-^{3}P_{0}$) and CO($J$=3$-$2) data were taken between 
December 1998 and March 1999 using the Mt. Fuji submillimeter-wave telescope. The diameter of the main 
reflector is 1.2 m, corresponding to a HPBW of $2'.2$ and $3'.0$ at 492 GHz and 346 GHz, 
respectively. The telescope is enclosed in a space frame radome whose transmission efficiency 
is 0.8 at 492 GHz and 0.9 at 346 GHz. The pointing of the telescope was checked and corrected 
by observing 346 GHz continuum emission from the Sun and the Moon every month, and the pointing accuracy 
has been maintained within $20''$(rms) during the observing run. We used a 346/492 GHz dual band 
SIS mixer receiver in our observations. Typical system temperatures including the atmospheric 
attenuation were 500 K (DSB) at 346 GHz 
and 1500 K (SSB) at 492 GHz. The backend is a 1024 channel acousto-optical spectrometer which 
has a total bandwidth of 900 MHz and an effective spectral resolution of 1.6 MHz. We split the 
spectrometer into two halves, each with 450 MHz bandwidth, to allow simultaneous observations of 
the CI and CO($J$=3-2) lines. Further details of the telescope will be described 
elsewhere \citep{sek99,mae99}.

We observed using position switching, where the off-source position was at 
($\alpha_{1950}$, $\delta_{1950}$) = (${\rm 05^{h} 28^{m} 46^{s}.5, -05^{\circ} 54' 28".0}$) for observations 
of the northern region and (${\rm 05^{h} 32^{m} 00^{s}.0, -07^{\circ} 18' 00".0}$) for the southern region, which 
were free of line emission to an rms noise level 
of 40 mK in the 1.6 MHz resolution. The intensity scale was calibrated using a chopper-wheel method. 
The moon efficiency including the radome loss is measured to be 0.75 at 346 GHz and 0.72 at 492 GHz. 
We will present intensities in the main-beam temperature scale ($T_{\rm MB}$) throughout this paper. 
The line intensities were checked every 4 hours by observing Orion-KL. The overall relative uncertainty 
in the final intensity scale is estimated to be within 20 \%. 
The zenith optical depth at 492 GHz ranged from 0.4 to 1.0 during the observations.

We have observed an $\sim$9 square degree area of the Orion A cloud 
with a grid spacing of $3'$. For most positions, the CI and CO($J$=3-2) lines were 
observed simultaneously. Furthermore we have taken additional CI data with a grid 
spacing of $1'.5$ for an $\sim$0.9 square degree region around Orion-KL and L1641N. 
In total, 4613 CI spectra and 3087 CO($J$=3-2) spectra were obtained. 
The on-source integration time ranged from 20 to 40 seconds per position and yielded typical 
rms noise temperatures of 0.5 K for CO($J$=3-2) and 0.6 K for CI. In this letter we concentrate on 
the global distributions of CI and CO($J$=3-2).

\section{Overall distribution of CI and CO($J$=3-2)}

Figure 1a shows the intensity map for CI, integrated between 3 km s$^{-1}$ and 13 km s$^{-1}$. 
CI emission is detected over almost the entire 
region of the Orion A cloud. The strongest CI emission is seen toward 
($\Delta\alpha$, $\Delta\delta$) = ($-40''$, $-220''$) from Orion-KL 
(${\rm 05^{h}32^{m}46^{s}.5, -05^{\circ}24'28''}$), where the peak temperature is 14.0 K, and slightly weaker 
toward Orion-KL. This trend was also seen in the higher resolution 
beam of \citet{wh95}. At Orion-KL the peak temperature is 9.1 K, the FWHM line width 
4.4 km s$^{-1}$, and the peak LSR velocity 9.4 km s$^{-1}$, which agree closely with the results 
reported with a similar beam size by \citet{ph81}. In the CI map, the $\int$-shaped 
filament reported by \citet{bal87} is clearly seen. At the southern end of the $\int$-shaped 
filament, a large dark cloud called L1641N can be identified, which is known to be a formation 
site of a low-mass cluster \citep{ho93}. The peak temperature of CI ranges up to 7 K around this 
region. From the south of L1641N, a filamentary structure continues to the southern end of the 
cloud with an almost constant width of about 4.4 pc. 
The left edge of this filament forms a straight line, and the filament is broken into a 
number of smaller clumps. The CI intensity decreases toward the south with 
$T_{\rm MB}\le$3 K and $\Delta v\sim$2.5 km s$^{-1}$, similar to values 
reported for HCL2 \citep{mae99}. 
The overall distribution of the CI emission closely resembles that of 
$^{13}$CO($J$=1-0) by \citet{bal87} with a similar ($1'.7$) beam to the CI observations.

Figure 1b shows the integrated intensity map for CO($J$=3-2). 
The emission peaks at Ori-KL where its $T_{\rm MB}$=67.8 K and the line width $\Delta v$= 
5.8 km s$^{-1}$. The line profile of CO($J$=3-2) shows wing emission originating 
from the molecular outflow, which is not seen in the CI spectra. The CO($J$=3-2) 
intensity drops sharply away from Ori-KL, and the $\int$-shaped filament is less clearly 
seen than in the CI map. In the central and southern parts of the cloud, the CO($J$=3-2) 
intensity distribution is rather featureless compared to that of CI. 
Although the large scale distribution of CO($J$=3-2) is similar to that of CI, the 
spatial extent is much larger. 
These features are probably due to a large optical depth in the CO($J$=3-2) line. 
Toward the southern part of the Orion A cloud, $T_{\rm MB}$ of CO($J$=3-2) 
is typically 3 K and $\Delta v\sim$3.0 km s$^{-1}$, similar to that of the CI 
line. 

\section{CI/CO($J$=3-2) intensity ratio}

Figure 2a shows a map of the integrated intensity ratio of CI/CO($J$=3-2).
The ratio shows a gradient from north to south. 
Around the Orion-KL region the ratio is as low as 0.10, increasing to 0.29 in 
L1641N, and 1.2 at the southern end of the cloud. 
The total intensity ratio for the Orion A cloud is evaluated to be 0.37. This value is slightly 
lower than the value for the Galactic plane, 0.57, reduced to the intensity ratio from the 
values observed by the COBE satellite 
\citep{wr91}. 

The CI optical depth has been suggested to be small or moderate ($\sim$3) 
for a wide range of UV fields and densities \citep{zm88,pl99}. 
By contrast the optical depth of CO($J$=3-2) is expected to be much larger 
than that of CI. The CI/CO($J$=3-2) ratio is sensitive to the 
optical depth of CI, if the CO($J$=3-2) line is saturated and the excitation temperatures 
for both lines are similar. 
The observed gradient suggests that $\tau$(CI) increases 
from the northern to the southern parts. 
The CI optical depth depends on the excitation temperature and on the
column density ($N$(CI)).  The gas kinetic temperature is known to have a
spatial gradient, from 60 K at Orion-KL to $\sim$15 K at the southern end of L1641
(Tatematsu \& Wilson 1998).  If we assume the LTE condition, the CI
optical depth increases by a factor of 6 from north to south with a fixed
column density.  Thus, the optical depth gradient is likely to reflect the
temperature gradient, although a gradient in the CI /CO abundance
ratio cannot be ruled out completely.

\section{Correlation between CI and $^{13}$CO($J$=1-0)}

Figure 2b shows a map of the integrated intensity ratio of CI/$^{13}$CO($J$=1-0), 
where the $^{13}$CO($J$=1-0) data were taken from \citet{bal87}. No systematic gradient 
is seen in this map. If we assume that the $^{13}$CO($J$=1-0) line is optically thin for 
the entire cloud, 
the CI/$^{13}$CO($J$=1-0) integrated intensity ratio approximately expresses the 
optical depth ratio $\tau$(CI)/$\tau$($^{13}$CO). Since the column density ratio 
$N$(CI)/$N$(CO) is proportional to the optical depth ratio at a given temperature, 
our result may suggest an almost uniform $N$(CI)/$N$(CO) ratio from north to 
south along the cloud regardless of the strength of the UV field. In order to confirm this, 
we derived the column densities of CI and CO under the LTE condition toward several 
representative positions as shown in Table 1, and find that the $N$(CI)/$N$(CO) 
ratio remains almost constant. However, the ratio along the 
ridge of the filament tends to be slightly lower than toward the edges. This trend 
is particularly clear in the $\int$-shaped filament, and similar to that reported by \citet{pl99} 
for much smaller regions toward W3, NGC2024, S140, and Cep A.

Figure 2b also suggests that the integrated intensity of CI correlates well with that 
of $^{13}$CO($J$=1-0). A correlation between CI and $^{13}$CO($J$=2-1) 
intensity has previously been suggested by \citet{tau95} and \citet{tate99} toward small portions of the Orion 
A cloud. Our results show that this correlation holds over an almost entire region of the Orion A cloud. 
In order to investigate this in detail, we prepared a correlation diagram by integrating the intensities 
over the 1 km s$^{-1}$ velocity width for the whole Orion A cloud (Figure 3a) and 
the southern region of the cloud (Figure 3b). 
The $^{13}$CO($J$=1-0) data were smoothed to a $3'$ grid for comparison with our CI data. 
The CI intensity has an offset at zero intensity 
from $^{13}$CO($J$=1-0), and increases almost linearly as the $^{13}$CO($J$=1-0) intensity increases. 
We least-square fitted the following equations; 
$\int T_{\rm MB}({\rm C_I}) dv = A \int T_{\rm MB}(^{13}{\rm CO}(J=1-0)) dv + B$, where 
we used the data above the 3$\sigma$ noise level for the $^{13}$CO($J$=1-0) in this analysis. 
The coefficients ($A$, $B$) are derived to be (0.55$\pm$0.02, 0.87$\pm$0.04 K km s$^{-1}$) and 
(0.46$\pm$0.03, 0.92$\pm$0.06 K km s$^{-1}$), and the correlation coefficients are 0.82 and 0.80 for the whole cloud and 
southern regions, respectively. It should be noted that the CI emission tends to saturate 
for the larger $^{13}$CO($J$=1-0) intensities as seen in Figure 3b. 

One explanation for these properties can be given in terms of a picture of a PDR. 
An almost identical CI distribution to that of $^{13}$CO($J$=1-0) could not easily be explained by 
homogeneous PDR models (e.g. Tielens and Hollenbach 1985). Taking this into account, 
we will assume that the cloud consists of a numerous small clumps,  
which are exposed to the external UV radiation. 
In a small clump with low visual extinction, all the CO is destroyed and 
only CI exists \citep{mo91}. This may be the reason why the offset in the CI intensity is seen. 
The H$_2$ column density of such a clump is 
estimated from the $B$ constants to be $\sim$1$\times 10^{21}$ cm$^{-2}$, where the 
CI abundance is assumed to be 10$^{-4}$ \citep{suz92}. This corresponds to a visual extinction $Av\sim$1, which agrees with 
the depth of the CI layers in PDR models (e.g. K\"oster et al 1994; Spaans 1996). 
As the clump size increases, CO can exist in the central part of the clump. 
For larger clumps, the size of the CO core increases, and CI 
exists only near the clump surface. Therefore the CI emission tends to saturate for 
larger $^{13}$CO($J$=1-0) intensities. 

If the above picture based on the PDR model is correct, the $N$(CI)/$N$(CO) ratio should 
depend on the size distribution of clumps as well as the UV field intensity. It is therefore 
curious that the $N$(CI)/$N$(CO) ratio shows no such systematic gradient from the northern  
to the southern part of the Orion A cloud. Considering this fact, the possibility that CI 
co-exists with CO in the deep interior of the cloud should also be considered seriously. Evolutionary models 
\citep{suz92} and chemical bi-stability models \citep{leb93} may be potential candidates. Further 
observations including the CI ($^3P_2-^3P_1$) line are necessary to characterize physical 
conditions of the CI emitting regions, which will be a key to solve the above problem. 

\acknowledgments

We would like to acknowledge John Bally for allowing us to use their $^{13}$CO($J$=1-0) data 
in digital form. We are grateful to Tomoharu Oka for valuable discussions, and to Glenn White for his critical reading of 
the manuscript. This study is supported by Grant-in-Aid 
from the Ministry of Education, Science, and Culture (Nos. 07CE2002 and 11304010).

\clearpage

\figcaption[]{Integrated intensity of (a) CI ($^{3}P_1$--$^{3}P_0$) and (b) CO($J$=3-2) observed toward the Orion A cloud, over 
the range 3 to 13 km s$^{-1}$. The solid lines enclose the observed region. 
The contour levels for CI are from 6 - 39 K km s$^{-1}$ 
with intervals of 3 K km s$^{-1}$. 
The contour levels for CO($J$=3-2) are 6, 12, 18, 24, 30, 39, 48, 60, 90, 120, 150, 210, 270 and 330 K km s$^{-1}$. 
Spectra observed at offset positions of (0, 0) and ($15'$, $-60'$) are also shown in each figure, where (0,0) corresponds to the 
central position ($\alpha$, $\delta$)$_{1950}$ = (${\rm 05^h 32^m 46^s.5, -05^{\circ} 24' 28''}$). 
\label{Fig. 1}}

\figcaption[]{(a) Ratio of $\int T_{\rm MB}({\rm C_I})dv$/$\int T_{\rm MB}({\rm CO({\it J}=3-2)})dv$. 
The contour levels range from 0.1 to 0.9 with intervals of 0.2. 
(b) Ratio of $\int T_{\rm MB}({\rm C_I})dv$/$\int T_{\rm MB}({\rm ^{13}CO({\it J}=1-0)})dv$. 
The contour levels range from 0.3 to 0.9 with intervals of 0.2. The $^{13}$CO($J$=1-0) data was obtained from Bally et al. (1987). 
The velocity range integrated was from 3 to 13 km s$^{-1}$ for all lines. \label{Fig. 2}}

\figcaption[]{The intensities of CI integrated over 1 km s$^{-1}$ bins running from 3 to 13 km s$^{-1}$ are plotted against those of 
$^{13}$CO($J$=1-0). The $^{13}$CO($J$=1-0) data is obtained from Bally et al. (1987). The dashed lines show 
3 $\sigma$ levels. The thick solid lines denote linear fits to the data. 
The data correspond to (a) the whole region of the Orion A cloud, (b) the L1641S region (below declination -7.4$^{\circ}$). 
\label{Fig. 3}}

\clearpage
{
\begin{table}
\scriptsize
\begin{center}
\caption{Typical line parameters of CI and CO($J$=3-2) and column densities of CI. \label{tbl-1}} 
\begin{tabular}{ccrrcrrcrrrrcrrcc}
\tableline\tableline
 &  & 
\multicolumn{2}{c}{$T_{\rm MB}$ } & & 
\multicolumn{2}{c}{$\Delta v$ }  & & 
\multicolumn{2}{c}{$\int T_{\rm MB} dv$} &$T_{\rm ex}$\tablenotemark{b} & $\tau$(CI)& &
$N$(CI) & $N$(CO)\tablenotemark{c} && $N$(CI)/$N$(CO) \\
Source& Position &
\multicolumn{2}{c}{K} & & 
\multicolumn{2}{c}{km s$^{-1}$}  & & 
\multicolumn{2}{c}{K km s$^{-1}$} & K&&&
cm$^{-2}$ & cm$^{-2}$  &&    \\
\cline{3-4} \cline{6-7} \cline{9-10} 
   & ($\Delta\alpha$, $\Delta\delta$)\tablenotemark{a}&
CI & CO  & &
CI & CO  & &
CI & CO  & &&& 
$\times$10$^{17}$&$\times$10$^{17}$ && \\
\tableline
Orion KL & (0, 0) & 
9.1 & 67.8 & &  4.4 & 5.8 & & 46 & 475 &65.0&0.2&& 6.2 & 123 && 0.05 \\
L1641-N & ($15'$, $-60'$) & 
7.2 & 15.0 & &  3.3 & 5.0 & & 25 & 85 &21.6&0.9& & 3.2 & 31 && 0.10  \\
L1641-C & ($51'$, $-102'$) & 
5.4 & 6.0 & &  2.2 & 3.7 & & 14.1 & 25 &15.9&1.5&& 2.1 & 9.8 && 0.21 \\
L1641-S4 & ($90'$, $-165'$) & 
3.6 &  3.8 & &  3.2 & 2.8 & & 13.8 & 11.4 &16.0&0.7&& 2.0 & 14.4 && 0.14 \\
\tableline

\tablenotetext{a}{Offsets are relative to the central position ($\alpha$, $\delta$)$_{1950}$ = 
(${\rm 05^h 32^m 46^s.5, -05^{\circ} 24' 28''}$).}
\tablenotetext{b}{Taken from \citet{nag98}. $T_{\rm ex}$ for Orion KL is taken from Fig. 5 and for other positions from 
Table 2.}
\tablenotetext{c}{The column density of CO is calculated from the $^{13}$CO($J$=1-0) data 
\citep{bal87} assuming that CO/$^{13}$CO is 60 \citep{la93}.}

\end{tabular}
\end{center}
\end{table}
}

\clearpage


\begin{thebibliography}{}

\bibitem[Bally et al.(1987)]{bal87} Bally, J., Langer, W. D., Stark, A. A., and Wilson, R. W. 1987, \apj, 312, L45
\bibitem[Genzel \& Stutzki(1989)]{ge89} Genzel, R., \& Stutzki, J. 1989, ARA\&A, 27, 41
\bibitem[Hodapp \& Deane(1993)]{ho93} Hodapp, K., \& Deane, J. 1993, \apjs, 88, 119
\bibitem[K\"oster et al.(1994)]{ko94} K\"oster, B., St\"orzer, H., Stutzki, J., and Sternberg, A. 1994, A\&A, 284, 545
\bibitem[Langer \& Penzias(1993)]{la93} Langer, W. D., \& Penzias, A. A. 1993, \apj, 408, 539
\bibitem[Le Bourlot et al.(1993)]{leb93} Le Bourlot, J., des For\^ets, G. P., and Roueff, E. 1993, \apj, 416, L87
\bibitem[Maezawa et al.(1999)]{mae99} Maezawa, H., et al. 1999, submitted
\bibitem[Minchin et al.(1994)]{min94} Minchin, N. R., White, G. J., Stutzki, J., and Krause, D. 1994, A\&A, 291, 250
\bibitem[Monteiro(1991)]{mo91} Monteiro, T. T. 1991, A\&A, 241, L5
\bibitem[Nagahama et al.(1998)]{nag98} Nagahama, T., Mizuno, A., Ogawa, H., and Fukui, Y. 1998, \aj, 116, 336
\bibitem[Phillips and Huggins(1981)]{ph81} Phillips, T. G., and Huggins, P. J. 1981, \apj, 251, 533
\bibitem[Plume et al.(1994)]{pl94} Plume, R., Jaffe, D. T., and Keene, J. 1994, \apj, 425, L49
\bibitem[Plume et al.(1999)]{pl99} Plume, R., Jaffe, D. T., Tatematsu, K., Evans II, N. J., and Keene, J. 1999, \apj, 512, 768
\bibitem[Sakamoto et al.(1994)]{sak94} Sakamoto, S., Hayashi, M., Hasegawa, T., Handa, T., and Oka, T. 1994, \apj, 425, 641
\bibitem[Sekimoto et al.(1999)]{sek99} Sekimoto, Y., et al. 1999, in preparation   
\bibitem[Suzuki et al.(1992)]{suz92} Suzuki, H., Yamamoto, S., Ohishi, M., Kaifu, N., Ishikawa, S., Hirahara, Y., and 
Takano, S. 1992, \apj, 392, 551
\bibitem[Spaans(1996)]{sp96} Spaans, M. 1996, A\&A, 307, 271
\bibitem[Tatematsu et al.(1993)]{tate93}Tatematsu, K., et al. 1993, \apj, 404, 643
\bibitem[Tatematsu et al.(1998)]{tate98} Tatematsu, K., Umemoto, T., Heyer, M. H., Hirano, N., Kameya, O., 
and Jaffe, D. T. 1998, \apjs, 118, 517
\bibitem[Tatematsu \& Wilson(1998)]{tw98} Tatematsu, K., \& Wilson, T. L. 1998, In the Orion Complex Revisited, A.S.P. Conference 
Series, ed. M. McCaughrean \& A. Burkert (San Francisco: Astronimical Society of the Pacific), in press
\bibitem[Tatematsu et al.(1999)]{tate99} Tatematsu, K., Jaffe, D. T., Plume, R., Evans II, N. J., and Keene, J. 1999, \apj, 
in press
\bibitem[Tauber et al.(1995)]{tau95} Tauber, J. A., Lis, D. C., Keene, J., Schilke, P., and B\"uttgenbach, T. H. 1995, A\&A, 297, 567
\bibitem[Tielens \& Hollenbach(1985)]{tie85} Tielens, A. G. G. M., \& Hollenbach, D. 1985, \apj, 291, 722
\bibitem[White \& Padman(1991)]{wh91} White, G. J., \& Padman, R. 1991, \nat, 354, 511
\bibitem[White \& Sandell(1995)]{wh95} White, G. J., \& Sandell, G. 1995, A\&A, 299, 179 
\bibitem[Wright et al.(1991)]{wr91} Wright, E. L., et al. 1991, \apj, 381, 200 
\bibitem[Zmuidzinas et al.(1988)]{zm88} Zmuidzinas, J., Betz, A. L., Boreiko, R. T., and Goldhaber, D. M. 1988, \apj, 335, 774
\end{thebibliography}
\end{document}